\documentclass[10pt,twocolumn,letterpaper]{article}

\usepackage[pagenumbers]{wacv} 

\usepackage{graphicx}
\usepackage{amsmath}
\usepackage{amssymb}
\usepackage{booktabs}
\usepackage{tikz}
\usepackage{multirow}
\usepackage{booktabs}

\usepackage{colortbl} 
\usepackage{xcolor} 
\definecolor{lightred}{rgb}{1.0, 0.94, 0.94}
\usepackage{amsmath}
\usepackage{algorithm}
\usepackage{algpseudocode}
\usepackage{makecell}
\usepackage{bm}
\usepackage{tikz}
\usepackage{amsmath}
\usetikzlibrary{positioning, shapes.geometric, fit}

\usepackage[pagebackref,breaklinks,colorlinks]{hyperref}

\usepackage[capitalize]{cleveref}
\crefname{section}{Sec.}{Secs.}
\Crefname{section}{Section}{Sections}
\Crefname{table}{Table}{Tables}
\crefname{table}{Tab.}{Tabs.}

\def\wacvPaperID{2528} 
\def\confName{WACV}
\def\confYear{2025}

\begin{document}

\title{Multi-Cohort Framework with Cohort-Aware Attention and Adversarial Mutual-Information Minimization for Whole Slide Image Classification}

\author{Sharon Peled, Yosef E. Maruvka, Moti Freiman\\ Technion – Israel Institute of Technology.\\ Haifa, Israel.\\
{\tt\small sharonpe@campus.technion.ac.il}
}
\maketitle

\begin{abstract}

Whole Slide Images (WSIs) are critical for various clinical applications, including histopathological analysis. However, current deep learning approaches in this field predominantly focus on individual tumor types, limiting model generalization and scalability.
This relatively narrow focus ultimately stems from the inherent heterogeneity in histopathology and the diverse morphological and molecular characteristics of different tumors.
To this end, we propose a novel approach for multi-cohort WSI analysis, designed to leverage the diversity of different tumor types.
We introduce a Cohort-Aware Attention module, enabling the capture of both shared and tumor-specific pathological patterns, enhancing cross-tumor generalization.
Furthermore, we construct an adversarial cohort regularization mechanism to minimize cohort-specific biases through mutual information minimization. 
Additionally, we develop a hierarchical sample balancing strategy to mitigate cohort imbalances and promote unbiased learning.
Together, these form a cohesive framework for unbiased multi-cohort WSI analysis.
Extensive experiments on a uniquely constructed multi-cancer dataset demonstrate significant improvements in generalization, providing a scalable solution for WSI classification across diverse cancer types. 
Our code for the experiments is publicly available at \url{<link>}.
\end{abstract}

\section{Introduction}
Whole Slide Images (WSIs) are high-resolution digital scans of tissue biospecimens, designed to capture detailed cellular and morphological patterns for histological examination \cite{ghaznavi2013digital, dimitriou2019deep}. WSIs play a key role in clinical diagnosis and prognosis, often regarded as ground truth in medical practice \cite{kumar2020whole, hanna2020whole, kumar2020whole, mariani2018interstitial}. 

\begin{figure}[ht]
\centering
\includegraphics[width=1\columnwidth]{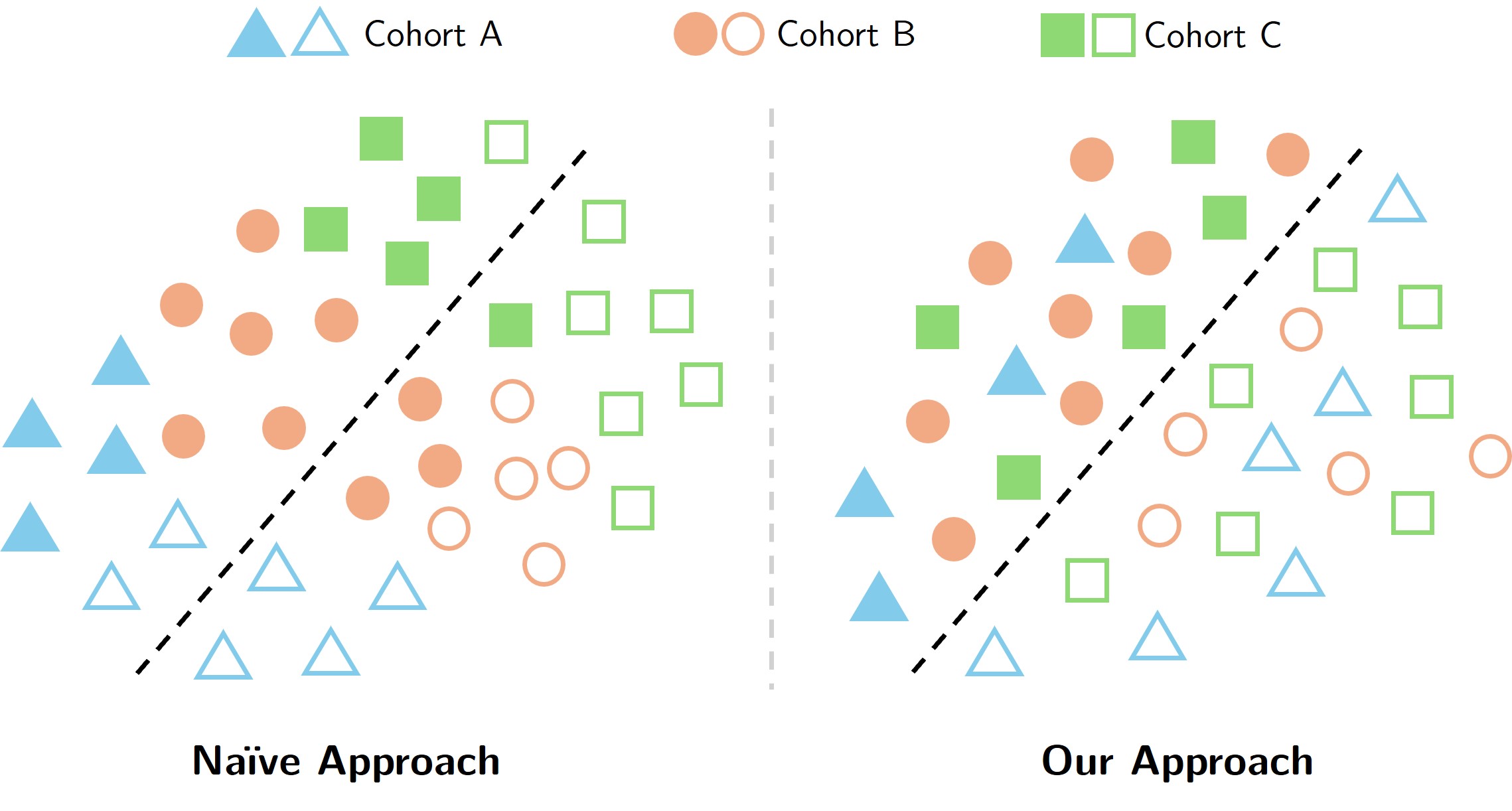}
\caption{Multi-cohort analysis paradigms. \textbf{Left}: Standard joint training, lacks mitigation of cohort biases. \textbf{Right}: Our framework, learning shared underlying mechanisms across cohorts.
}
\label{concept}
\end{figure}

In recent years, there has been a tremendous amount of work dedicated to applying deep learning to WSIs \cite{banerji2022deep, srinidhi2021deep, abdel2023comprehensive, bilal2021development, chan2023histopathology}, where deep learning has showcased its potential to impact clinical practices in pathology \cite{wu2022recent}. 
Due to the immense size of WSIs, a widely adopted approach is to decompose these large images into smaller, processable patches \cite{bilal2021development, zhang2022dtfd, kather2019deep, kumar2020whole}. Given that annotations are often provided at the slide level, histopathological image analysis is commonly formulated as a Multiple Instance Learning (MIL) task \cite{zhang2022dtfd, shao2021transmil, li2021dual, bontempo2023graph}, where each slide is treated as a collection of instances (patches).

While significant strides have been made in applying deep learning to WSIs, the vast majority of studies have concentrated on analyzing individual tumor types. This relatively narrow focus, introduces several limitations. First, the reliance on single-tumor datasets limits the amount of available data, often resulting in models that are data-hungry yet constrained by small sample sizes, which hinders their ability to generalize effectively. Additionally, the rich diversity of morphological patterns within each slide further increases the risk of overfitting, particularly when considering datasets limited to a single cancer type, which often composed of a few hundred samples.
Moreover, from a practical standpoint, maintaining separate models for each tumor type in clinical practice can be both computationally expensive and logistically challenging. The need for constant updates and management of multiple models could complicate their adoption into routine diagnostics.

Incorporating various cancer types within a unified framework could overcome these limitations by broadening the model's exposure to diverse pathological patterns, enabling it to learn richer representations. This approach could not only enhance the model's generalization capabilities but also potentially provide a deeper understanding of cancer’s underlying mechanisms through the discovery of both common and unique cancer markers. See \cref{concept} for an illustration. Finally, such a framework would simplify maintenance in clinical practice, which can increase confidence in the system and facilitate easier adoption \cite{park2021aggregation}. 

However, such integration introduces a new set of challenges, primarily due to the inherent heterogeneity in histopathology and the varied phenotypes across different tumors. One of the key challenges is the alignment of diverse pathological features across multiple cohorts, which complicates the training and evaluation of the model.
Additionally, a multi-source dataset may exacerbate existing imbalances, which can skew the model’s learning process, leading to biased predictions and reduced performance.

To this end, we propose a new framework designed to leverage the diversity of tumors in a multi-cohort WSI analysis. Our approach features a Cohort-Aware Attention Encoder, engineered to extract features unique to each tumor type while synthesizing shared pathological patterns across cohorts. This dual capability enhances the model’s ability to generalize across diverse cancer types while ensuring that both commonalities and cohort-specific nuances are effectively captured during training.
We also introduce an adversarial cohort regularizer that minimizes the mutual information (MI) between MIL representations and cohort associations, guiding the MIL representations to converge toward universally meaningful features and mitigating cohort biases that could otherwise lead to skewed predictions. Additionally, we propose a novel sample balancing strategy to address the hierarchical imbalances inherent in a multi-cohort environment, ensuring that the optimization process aligns with the downstream task.
Together, these components form a cohesive framework for comprehensive and unbiased multi-cohort WSI analysis.
The contributions of this paper are summarized as follows:
\begin{enumerate}
    \item \textbf{Cross-tumor learning framework}: We propose a novel framework that integrates multiple cancer types, significantly improving generalization over existing state-of-the-art methods by enabling the discovery of both shared and tumor-specific pathological patterns.

    \item \textbf{Cohort-Aware Attention Module}: We introduce an innovative Cohort-Aware VisionTransformer (CAViT) as our tile encoder, enabling the capture of cohort-specific features while synthesizing shared pathological patterns across cohorts.

    \item \textbf{Adversarial cohort regularization}: We implement an adversarial regularizer that minimizes mutual information between cohort labels and MIL representations, mitigating cohort biases and guiding the model toward universally meaningful features.
    
    \item \textbf{Hierarchical sample balancing}: We present a novel hierarchical sample balancing strategy to address cohort imbalances, ensuring unbiased learning aligned with the downstream task.
    
    \item \textbf{Cross-tumor benchmark dataset}: We establish a benchmark dataset that integrates multiple cancer types, providing a foundation for evaluating multi-cohort learning strategies in digital pathology.
    
\end{enumerate}


\section{Related Work}
\paragraph{\textbf{Multi-Cohort Analysis in Histopathology}}
While cross-tumor analyses, such as pan-cancer studies \cite{kather2020pan, fu2020pan}, have been conducted, these efforts primarily focus on identifying common mutations and genomic variations rather than integrating distinct cancer types into a cohesive learning framework. The exploration of multi-cohort learning in histopathology remains relatively limited.
For instance, \cite{noorbakhsh2020deep} showed that although 'universal' models tend to underperform relative to tumor-specific models, identifying cancer types with similar morphological features can lead to notable performance improvements.
Subsequently, \cite{park2021aggregation} proposed a strategy to form 'super-cohorts', which are groups of cancer types whose training conjointly can improve model performance. 
Although these approaches highlight the potential of combining different types of cancer, a holistic framework to effectively capitalize on the diversity of tumors is still a necessity.

\textbf{Multi-Dataset in Deep Learning}
In recent years, significant progress has been made in applying deep learning to multi-dataset (MD) scenarios \cite{zhou2022simple, chen2024versatile, hu2021unit}. However, much of this progress has focused on multi-dataset multi-task (MD-MT) learning, which aims to address multiple tasks simultaneously using the same input data \cite{hu2021unit, xu2023multi, zhao2023multi}.
These studies naturally emphasize the interaction between tasks, leveraging the inherent correlations among related tasks to achieve reciprocal benefits and improve overall performance.
For studies that focus on multi-dataset interactions, the primary approach has often been to combine losses \cite{chen2024versatile, zhou2022simple,hu2021unit } or fuse label spaces \cite{zhou2022simple, chen2023scaledet}, thereby facilitating interactions between different datasets. 
Furthermore, existing approaches for these learning paradigms (MT-MD or MD) typically require large datasets with millions of examples to effectively learn the joint dataset space \cite{hu2021unit, xu2023multi, zhao2023multi, zhou2022simple, chen2023scaledet}.
Our approach directly models the interactions between datasets during joint training, enabling the facilitation of even small datasets composed of only a few hundred samples.

\textbf{Cross-Attention}
Self-attention has been extensively adapted from its original mechanism \cite{vaswani2017attention} to facilitate interactions between multiple entities \cite{hassanin2024visual, lu2019vilbert}. One prominent adaptation is co-attention \cite{lu2019vilbert}, which enables interactions between different modalities during joint training by using the query matrix to facilitate cross-entity interactions. This key observation - that the query matrix can be used to form cross interactions - has been widely explored and applied in various contexts \cite{xu2023multi, lee2022task, lopes2023cross, hu2021unit, bhattacharjee2022mult}. 
Typically, cross-attention adaptations are designed for tasks involving multiple entities, such as multi-modal learning, where different modalities are processed and cross-attention can enhance overall performance \cite{lu2019vilbert, kim2023cross, mercea2022audio, wei2020multi}, or multi-task learning, where cross-attention facilitates interaction among different tasks during training \cite{xu2023multi, lee2022task, lopes2023cross, hu2021unit}. 
Direct cross-attention is not feasible in multi-cohort, uni-modal datasets due to the lack of inherent entity interactions. Our Cohort-Aware approach introduces a novel attention mechanism designed to overcome this limitation. Unlike prior methods, our approach leverages token-wise attention through a learnable Query-Attention (QA) component, dynamically integrating dataset-wide and cohort-specific queries. This allows each token to attend to either global patterns or cohort-specific features based on contextual relevance, ultimately improving generalization across cohorts.

\textbf{Hierarchical Imbalance}
Traditional class balancing techniques \cite{johnson2019survey, wang2016training, huang2016learning, reshma2021natural, khened2021generalized}, often fall short in multi-dataset setups with hierarchical structures. To address these challenges, several approaches have been proposed, such as class-aware sampling or hierarchical sampling \cite{zhou2022simple, hu2021unit, chen2024versatile}.
While effective in many cases, we find that sampling can be less reliable with small datasets, leading to skewed training batches and suboptimal performance. 
To overcome this, we propose a novel hierarchical sample weighting strategy that adapts to different levels of data granularity in multi-cohort WSI analysis. Our approach enforces a balanced loss, ensuring more robust performance across datasets with significant disparities.

\textbf{Adversarial Learning}
Adversarial learning, originally introduced in the context of Generative Adversarial Networks (GANs), has since been widely adopted across various applications. For example, in domain adaptation adversarial loss has proven effective in aligning feature representations between different domains \cite{tzeng2017adversarial,chen2020adversarial,tang2020discriminative}, and within a multi-task learning framework, adversarial strategies have been utilized to disentangle and refine task-specific representations \cite{liu2017adversarial,liu2018multi,zhou2020task}.
Our approach incorporates an adversarial cohort regularizer that controls the mutual information between MIL representations and cohort associations during training. This process enables the alignment of diverse pathological representations across multiple cohorts while effectively mitigating cohort-specific biases.

\begin{figure}[ht]
\centering
\includegraphics[width=0.85\columnwidth]{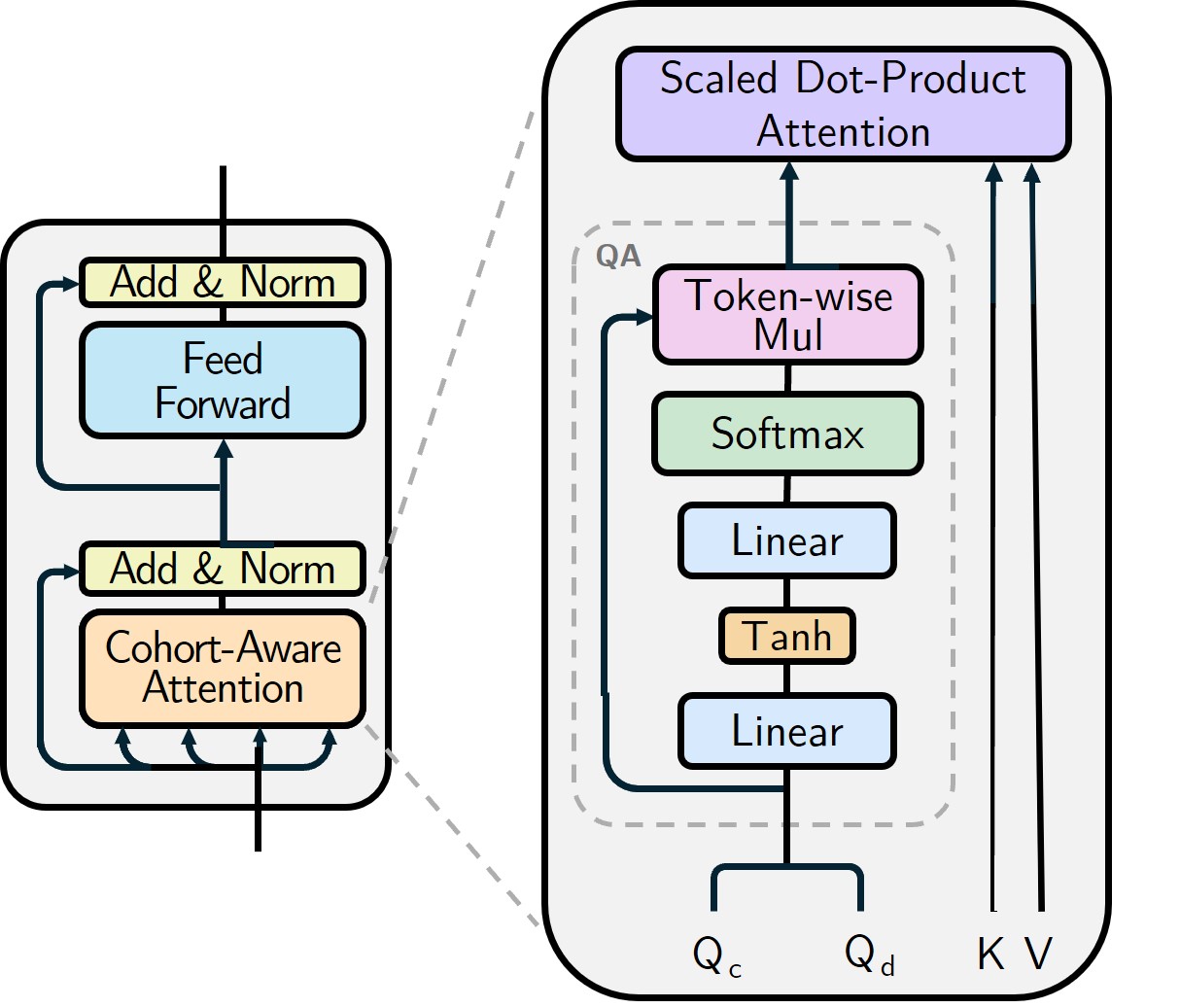}
\caption{Cohort-aware attention module integrated into a VisionTransformer \cite{dosovitskiy2020image} block. The dataset-wide query \( Q_d \) and cohort-specific query \( Q_c \) are processed through a token Query-Attention (QA) component to form \( Q_{ca} \), which is then fed into the scaled dot-product attention.
}
\label{cohort_aware}
\end{figure}

\section{Methodology}

\subsection{Preliminaries}


Let \( S = \{s_1, s_2, \dots, s_N\} \) represent a set of WSIs, where each slide \( s_i \) is associated with a cohort \( c_i \in \{1, 2, \dots, k\} \) and a label \( y_i \in \{1, 2, \dots, m\} \). Each slide is divided into tiles, which are processed by a pretrained encoder \( E(\cdot) \). Each slide \( s_i \) is represented by its collection of encoded tile features \( \{f_{ij}\}_{j=1}^{n_i} \in \mathbb{R}^d \). These features are then passed through a MIL model \( M(\cdot) \), which aggregates them to a final prediction \( \hat{y}_i \) for the slide.

\subsection{Cohort-Aware Attention}
\label{cavit}
In this paper, we present a novel Cohort-Aware Attention (CAA) module designed to address the challenges of multi-cohort WSI analysis. This module is capable of capturing cohort-specific variations and commonalities, thereby mitigating cohort biases and enhancing model performance across the entire dataset. While the CAA module can be applied in various ways, in this work, we deploy it within a VisionTransformer (ViT) \cite{dosovitskiy2020image}, creating a Cohort-Aware VisionTransformer (CAViT) as our tile encoder.

The key innovation of the CAA module lies in its ability to facilitate interactions between data cohorts during a joint training. Unlike conventional cross-attention, typically applied in multi-modal or multi-task setups with naturally interacting entities, our approach tackles the challenge of a multi-cohort, uni-modal dataset where direct cross-attention is not feasible. The CAA module enables cross-cohort interactions by engaging different components of the self-attention mechanism during training, capturing both shared and cohort-specific patterns.\\

Given an input matrix \( X \in \mathbb{R}^{n \times d} \), where \( n \) denotes the length of the patch sequence and \( d \) represents the feature dimension, along with a cohort identifier \( c \in \{1, 2, \dots, k\} \), the CAA employs distinct projection matrices for cohort-specific queries \( W_{Q_c} \in \mathbb{R}^{d \times d_k} \) and a shared dataset-wide query matrix \( W_{Q_d} \in \mathbb{R}^{d \times d_k} \), while utilizing common key and value matrices \( W_K, W_V \in \mathbb{R}^{d \times d_k} \) across all cohorts.
The queries, keys, and values are computed as:
\begin{align}
Q_d = XW_{Q_d}, \quad Q_c = XW_{Q_c},
\end{align}
\begin{align}
K = XW_K, \quad V = XW_V
\end{align}
The cohort-aware query \( Q_{ca} \) is then constructed by dynamically integrating the dataset-wide query \( Q_d \) with the cohort-specific query \( Q_c \).
This integration is facilitated by a learnable Query-Attention (QA) component, which dynamically maps each individual token query \( q_i \in \mathbb{R}^{1 \times d_k}\) (where \( q_i \) is the \( i \)-th query vector from the query matrix) to its corresponding attention weights:

\begin{align}
\alpha_d = \text{QA}\left(Q_d\right), \quad \alpha_c = \text{QA}\left(Q_c\right)
\end{align}
where \(\alpha_d, \alpha_c \in \mathbb{R}^{n \times 1}\) is the attention weights for the dataset query and cohort query of each token, respectively. 
The QA component is formulated as a non-linear fully-connected network, which projects each query token \( q_{i_d} \) or \( q_{i_c} \) to its corresponding attention weight. See \cref{cohort_aware} for illustration.
The cohort-aware query \( Q_{ca} \) is then obtained as the element-wise product of the attention weights with their respective queries:

\begin{align}
Q_{ca} = \alpha_d \odot Q_d + \alpha_c \odot Q_c
\end{align}
The Cohort-Aware (CA) attention weights are computed using Scaled Dot-Product Attention \cite{vaswani2017attention, dosovitskiy2020image} as follows:

\begin{equation}
     \text{Attention}(Q_{ca}, K, V) =\text{softmax}\left(\frac{Q_{ca}K^T}{\sqrt{d_k}}\right)V
    \end{equation}
The \( Q_{ca} \) matrix adapts dynamically, allowing inputs to attend to either global patterns via \( Q_d \) or cohort-specific features via \( Q_c \), depending on the contextual importance.
It is crucial that each \( W_{Q_c} \) is updated exclusively with gradients derived from samples from cohort \( c \).
This selective gradient update mechanism allows \( W_{Q_c} \) to be precisely tuned to the distinct characteristics of cohort \(c\).

\subsubsection{Cohort-Aware VisionTransformer}
To create the Cohort-Aware VisionTransformer (CAViT), we modify the standard VisionTransformer \cite{dosovitskiy2020image} by replacing the Multihead Self-Attention (MHA) module \cite{vaswani2017attention} with our Multihead Cohort-Aware Attention (MCAA) module. In this architecture, the MCAA block processes both the input image \( x \) and cohort identifier \( c \) to generate attention scores that are attuned to cohort-specific characteristics while simultaneously learning global features. The output of the MCAA layer is then passed through a residual connection and layer normalization (Add \& Norm). This is followed by an fc layer and an additional Add \& Norm layer. The result is then propagated to the subsequent block in the CAViT architecture. See \cref{cohort_aware} for an illustration.

\begin{figure}[ht]
\centering
\includegraphics[width=0.75\columnwidth]{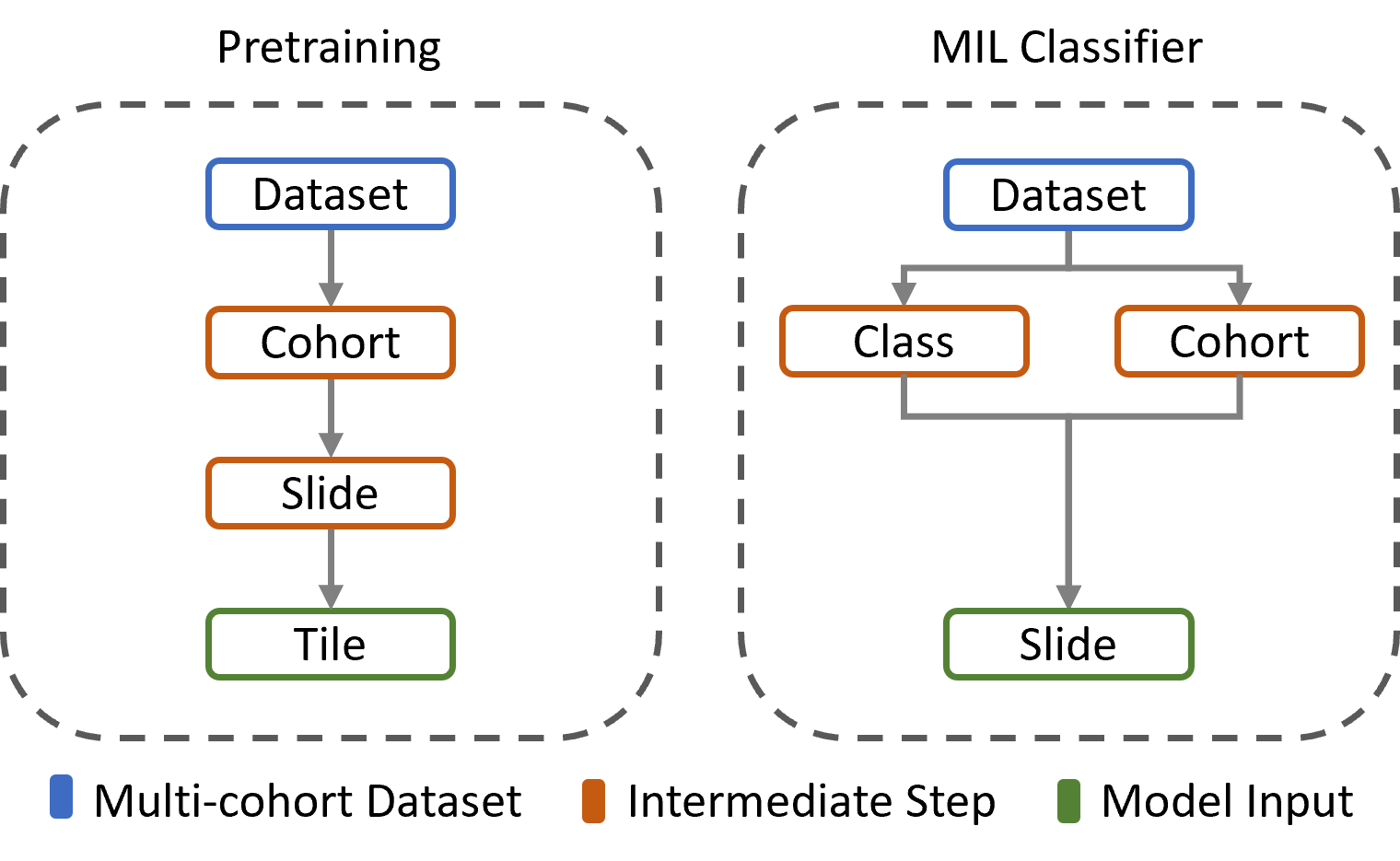}
\caption{Hierarchical data structure of multi-cohort WSI analysis during pretraining and MIL phases. The pretraining phase balances cohorts, slides, and tiles, ensuring proportional representation at each level. The MIL phase further balances cohort-class combinations and uniformly distributes weights among slides. This strategy mitigates imbalances, promoting effective and unbiased model learning.}
\label{task_balance}
\end{figure}

\subsection{Adversarial Cohort-Regularized Training for Multi-Cohort Learning}
In multi-cohort WSI analysis, a critical challenge arises from cohort-specific biases embedded within tile representations. Due to the amount of information each WSI encapsulates, the learning of undesired patterns is an inevitable risk.
Our key observation reveals that cohort associations can be predicted with high accuracy directly from MIL representations, regardless of whether a cohort-aware encoder is used. Even with a simple linear classifier, the predictive accuracy is considerably higher than a random classifier\footnote{AUC exceeds 80\% when predicting cohort association directly from MIL representations}, indicating that the model retains undesired cohort-specific information.
This retention risks overfitting, particularly in scenarios where cohorts exhibit different class imbalances, leading to biased predictions that compromise the model's generalization. 

To mitigate this, we seek to minimize the mutual information (MI) between MIL representations and cohort associations, thereby promoting the learning of unbiased representations. However, directly computing MI for continuous high-dimensional random variables is intractable. 
To address this issue, the Mutual Information Neural Estimation (MINE) algorithm \cite{belghazi2018mutual} was introduced. 
In our framework, we utilize a more recent advancement in MI estimation - the Smoothed Mutual Information "Lower Bound" Estimator (SMILE) \cite{song2019understanding} - which improves upon MINE by reducing variance.

\subsubsection{The Adversarial Network}
As previously defined, the Multiple Instance Learning (MIL) model \( M(\cdot) \) processes tile embeddings \( \{f_{ij}\}_{j=1}^{n_i} \) for each slide \( s_i \), producing a final prediction score \( \hat{y}_i = M(\{f_{ij}\}_{j=1}^{n_i}) \). To clearly separate the components within the MIL model, we denote the MIL aggregator function as \( A(\cdot) \), which aggregates the tile-level features into a slide-level representation:

\begin{align}
\label{eq:mil_agg}
z_i = A(\{f_{ij}\}_{j=1}^{n_i})
\end{align}

The slide-level representation \( z_i \) is then passed through the MIL head \( H(\cdot) \), to generate the final prediction:

\begin{align}
\label{eq:mil_head}
\hat{y}_i = H(z_i)
\end{align}

Our adversarial network leverages the \( I_{\text{SMILE}} \) estimator \cite{song2019understanding} to estimate the mutual information (MI) between the MIL representations \( z_i \in Z \) and the one-hot cohort association vectors \( c_i \in C \) during training. 

\( I_{\text{SMILE}} \) estimator utilizes a score network \( T_\theta \) \cite{belghazi2018mutual, song2019understanding} which designed to estimate the log-density ratio between the joint distribution \( P_{Z,C}(z,c) \) and the product of the marginal distributions \( P_Z(z)P_C(c) \). Our \( T_\theta \) consists of three non-linear transformations. First, the slide-level representation \( z_i \) and the cohort association \( c_i \) are independently mapped to hidden representations \( \mathbf{h}_{z_i} \), \( \mathbf{h}_{c_i} \in \mathbb{R}^h \), respectively. These hidden representations are then concatenated and further processed by a final transformation to produce the score \( T_\theta(z_i, c_i) \).

Given an input batch \(\{z_i\}_{i=1}^{B}\) and \(\{c_i\}_{i=1}^{B}\), positive pairs \((z_i, c_i)\) and negative pairs \((z_i, c_j)_{i \neq j}\) are constructed. The adversarial network is trained to maximize the \( I_{\text{SMILE}}(T_\theta, \tau) \) estimator:

\begin{align}
\label{eq:smile}
\mathbb{E}_{P_{Z,C}} [T_\theta(z, c)] - \log \mathbb{E}_{P_ZP_C}[\text{clip}(e^{T_\theta(z,c)}, e^{-\tau}, e^{\tau})]
\end{align}

where \( \text{clip}(v, l, u) = \max(\min(v, u), l) \), and \( \tau \) is a hyperparameter.

\subsubsection{Adversarial Training}

For a batch of tile embeddings \(\{x_i\}_{i=1}^{B} \), the slide-level representations \( z_i \) are generated using the aggregator \( A(\cdot) \) as defined in \cref{eq:mil_agg}. 
Positive and negative pairs are constructed from \(\{z_i\}_{i=1}^{B}\) and \(\{c_i\}_{i=1}^{B}\), and a backward pass is performed to update the adversarial network \( I_{\text{SMILE}} \), as outlined in \cref{eq:smile}.
Following this, an MI estimation loss \(-I_{\text{SMILE}}(\{z_i\}_{i=1}^{B}, \{c_i\}_{i=1}^{B})\) is calculated with the adversarial network now updated and frozen.
The MIL head \( H(\cdot) \) then generates the final prediction scores, as shown in \cref{eq:mil_head}.
The overall training objective is defined as:
\[
\mathcal{L} = \mathcal{L}_{\text{MIL}} + \lambda \mathcal{L}_{\text{MI}}
\]
where \(\mathcal{L}_{\text{MIL}}\) is the original MIL loss, \(\mathcal{L}_{\text{MI}}\) is the mutual information loss, and \( \lambda \) is a hyperparameter.
The procedure is summarized in \cref{algo}.

This adversarial framework compels the MIL model to produce unbiased representations, thus promoting the learning of more generalizable features across different cohorts.

\begin{algorithm}[ht]
\caption{Adversarial Cohort-Regularized Training}
\begin{algorithmic}[1]
\State Initialize MIL model and \(I\) MI estimator parameters.
\For{each mini-batch $\{(x_i, y_i, c_i)\}_{i=1}^{B}$}
    \State Compute slide-level representations $z_i = A(x_i)$.
    \State Estimate MI between \(\{z_i\}_{i=1}^{B}\) and \(\{c_i\}_{i=1}^{B}\) using \(I\).
    \State Update \(I\) via backpropagation.
    \State Freeze \(I\), re-estimate \(\mathcal{L}_{\text{MI}}\).
    \State Compute prediction scores $\hat{y}_i = H(z_i)$.
    \State Update MIL model by minimizing: 
    \[
    \mathcal{L} = \mathcal{L}_{\text{MIL}} + \lambda \mathcal{L}_{\text{MI}}.
    \]
\EndFor
\end{algorithmic}
\label{algo}
\end{algorithm}

\begin{table*}[ht]
\centering
\begin{tabular}{lcccccccc}
\toprule
\multirow{2}{*}{Method} & \multicolumn{2}{c}{All} & \multicolumn{2}{c}{CRC} & \multicolumn{2}{c}{STAD} & \multicolumn{2}{c}{UCEC} \\
\cmidrule(lr){2-3} \cmidrule(lr){4-5} \cmidrule(lr){6-7} \cmidrule(lr){8-9}
& AUC & B-Acc & AUC & B-Acc  & AUC & B-Acc & AUC & B-Acc \\
\midrule
Mean-pool & $71.53\textsubscript{ ±1.9}$ & $56.82\textsubscript{ ±1.9}$ & $78.58\textsubscript{ ±4.0}$ & $57.63\textsubscript{ ±1.5}$ & $67.23\textsubscript{ ±2.3}$ & $53.24\textsubscript{ ±2.6}$ & $58.08\textsubscript{ ±1.4}$ & $55.28\textsubscript{ ±2.4}$ \\
Mean-max & $68.94\textsubscript{ ±2.0}$ & $54.43\textsubscript{ ±1.8}$ & $68.04\textsubscript{ ±1.3}$ & $52.46\textsubscript{ ±0.5}$ & $65.74\textsubscript{ ±3.9}$ & $50.03\textsubscript{ ±0.5}$ & $64.82\textsubscript{ ±2.1}$ & $57.95\textsubscript{ ±3.1}$ \\
AB-MIL \cite{ilse2018attention} & $76.88\textsubscript{ ±2.1}$ & $62.18\textsubscript{ ±1.8}$ & $82.78\textsubscript{ ±4.1}$ & $60.87\textsubscript{ ±1.6}$ & $72.39\textsubscript{ ±2.7}$ & $62.42\textsubscript{ ±2.6}$ & $64.65\textsubscript{ ±1.3}$ & $60.15\textsubscript{ ±2.1}$ \\
GCN-MIL \cite{zhao2020predicting} & $78.00\textsubscript{ ±1.6}$ & $66.15\textsubscript{ ±1.2}$ & $77.02\textsubscript{ ±5.9}$ & $73.52\textsubscript{ ±4.4}$ & $81.17\textsubscript{ ±1.9}$ & $69.64\textsubscript{ ±3.0}$ & $66.15\textsubscript{ ±3.0}$ & $58.96\textsubscript{ ±1.9}$ \\
SparseConvMIL \cite{lerousseau2021sparseconvmil} & $76.89\textsubscript{ ±1.9}$ & $64.91\textsubscript{ ±2.3}$ & $84.34\textsubscript{ ±4.1}$ & $71.34\textsubscript{ ±4.8}$ & $80.05\textsubscript{ ±2.3}$ & $65.04\textsubscript{ ±3.4}$ & $65.91\textsubscript{ ±2.3}$ & $60.39\textsubscript{ ±1.9}$ \\
TransMIL \cite{shao2021transmil} & $74.69\textsubscript{ ±2.3}$ & $62.61\textsubscript{ ±1.6}$ & $80.33\textsubscript{ ±3.4}$ & $68.09\textsubscript{ ±1.9}$ & $79.47\textsubscript{ ±3.0}$ & $63.78\textsubscript{ ±1.9}$ & $62.61\textsubscript{ ±1.6}$ & $58.55\textsubscript{ ±0.9}$ \\
\midrule
DTFD-MIL \cite{zhang2022dtfd} & $78.44\textsubscript{ ±3.2}$ & $63.89\textsubscript{ ±1.5}$ & $82.09\textsubscript{ ±6.0}$ & $70.03\textsubscript{ ±1.5}$ & $74.83\textsubscript{ ±3.8}$ & $66.56\textsubscript{ ±1.5}$ & $63.53\textsubscript{ ±2.4}$ & $60.32\textsubscript{ ±1.5}$ \\
\rowcolor{lightred}
Ours (DTFD-MIL) & $79.50\textsubscript{ ±1.7}$ & $66.05\textsubscript{ ±1.3}$ & $\textbf{87.94\textsubscript{ ±2.4}}$ & $\textbf{76.73\textsubscript{ ±2.3}}$ & $\textbf{82.05\textsubscript{ ±2.3}}$ & $66.82\textsubscript{ ±2.3}$ & $67.04\textsubscript{ ±1.6}$ & $60.06\textsubscript{ ±2.1}$ \\
\midrule
MHA-MIL \cite{vaswani2017attention} & $77.19\textsubscript{ ±2.5}$ & $65.61\textsubscript{ ±2.0}$ & $84.04\textsubscript{ ±2.7}$ & $69.28\textsubscript{ ±2.7}$ & $80.80\textsubscript{ ±3.4}$ & $70.91\textsubscript{ ±3.9}$ & $66.10\textsubscript{ ±4.2}$ & $60.17\textsubscript{ ±4.1}$ \\
\rowcolor{lightred}
Ours (MHA-MIL) & $\textbf{80.60\textsubscript{ ±1.6}}$ & $\textbf{69.93\textsubscript{ ±1.4}}$ & $85.08\textsubscript{ ±3.1}$ & $75.35\textsubscript{ ±4.0}$ & $81.69\textsubscript{ ±2.2}$ & $\textbf{71.11\textsubscript{ ±1.3}}$ & $\textbf{69.34\textsubscript{ ±1.4}}$ & $\textbf{61.64\textsubscript{ ±2.0}}$ \\
\bottomrule
\end{tabular}
\caption{Comparative analysis of Area Under the Curve (AUC) and Balanced Accuracy (B-Acc) scores for the MSS/MSI task across CRC, STAD, and UCEC cohorts, along with the overall performance (All). The highest performance scores are highlighted in bold.}
\label{msi_compare}
\end{table*}

\subsection{Imbalance Mitigation}
Aggregating data from multiple cohorts increases training diversity but it may introduce imbalances at cohort, class, slide, or tile levels. If not properly addressed, these imbalances can skew the learning process. Existing methods, which often rely on sampling techniques \cite{zhou2022simple, hu2021unit, chen2024versatile}, struggle to effectively mitigate these imbalances in smaller datasets, where sampling can skew the composition of training batches, leading to suboptimal model performance.

To this end, we propose a new hierarchical balancing strategy that ensures equitable representation across multiple levels of data granularity inherent in a multi-cohort WSI analysis. See \cref{task_balance} for illustration.

In the pretraining phase, we balance weights across cohorts, slides, and tiles, ensuring equal contribution at each hierarchical level. During MIL training, we balance cohort-class combinations and distribute weights equally among slides within each combination, preventing any single cohort or class from dominating training.

WSI cohorts often display significant disparities in tile counts or class distributions, which may cause certain samples to be disproportionately weighted. Therefore, we clip sample weights within two standard deviations of the mean and apply batch normalization during training. 

This approach not only minimizes over- or under-representation but also acts as a regularization technique, dynamically adjusting sample weights based on batch composition during training.

\begin{table}
\centering
\small
\begin{tabular}{c@{\hspace{6pt}}cc@{\hspace{6pt}}cc}
\toprule
 & \multicolumn{2}{c}{MSS / MSI} & \multicolumn{2}{c}{GS / CIN} \\
\cmidrule(lr){2-3} \cmidrule(lr){4-5}
 & MSS & MSI & GS & CIN \\
\midrule
CRC  & 534 (86\%)  & 87 (14\%)  & 54 (14\%)  & 326 (86\%)  \\
STAD & 364 (82\%)  & 78 (18\%)  & 55 (20\%)  & 222 (80\%)  \\
UCEC & 382 (69\%)  & 175 (31\%) & 158 (49\%) & 166 (51\%) \\
\midrule
Total & 1344 (82\%) & 341 (18\%) & 268 (27\%) & 770 (73\%) \\
\bottomrule
\end{tabular}
\caption{Slide counts per cohort per subtype. Note the differences in sizes and class imbalances exhibited by each cohort.}
\label{slide_counts}
\end{table}

\begin{table*}[ht]
\centering
\begin{tabular}{lcccccccc}
\toprule
\multirow{2}{*}{Method} & \multicolumn{2}{c}{All} & \multicolumn{2}{c}{CRC} & \multicolumn{2}{c}{STAD} & \multicolumn{2}{c}{UCEC} \\
\cmidrule(lr){2-3} \cmidrule(lr){4-5} \cmidrule(lr){6-7} \cmidrule(lr){8-9}
& AUC & B-Acc & AUC & B-Acc  & AUC & B-Acc & AUC & B-Acc \\
\midrule
Mean-pool & $78.99\textsubscript{ ±1.6}$ & $71.72\textsubscript{ ±1.9}$ & $78.62\textsubscript{ ±1.5}$ & $66.63\textsubscript{ ±2.7}$ & $69.00\textsubscript{ ±2.2}$ & $61.20\textsubscript{ ±3.8}$ & $82.56\textsubscript{ ±1.4}$ & $85.72\textsubscript{ ±1.4}$ \\
Mean-max & $81.49\textsubscript{ ±1.5}$ & $75.40\textsubscript{ ±0.6}$ & $79.28\textsubscript{ ±2.0}$ & $66.81\textsubscript{ ±2.9}$ & $79.90\textsubscript{ ±2.4}$ & $72.43\textsubscript{ ±2.7}$ & $86.95\textsubscript{ ±2.3}$ & $79.79\textsubscript{ ±1.9}$ \\
AB-MIL \cite{ilse2018attention} & $84.00\textsubscript{ ±2.0}$ & $77.58\textsubscript{ ±1.5}$ & $79.15\textsubscript{ ±1.8}$ & $69.46\textsubscript{ ±2.2}$ & $79.31\textsubscript{ ±3.3}$ & $73.97\textsubscript{ ±2.8}$ & $89.72\textsubscript{ ±2.4}$ & $83.65\textsubscript{ ±1.9}$ \\
GCN-MIL \cite{zhao2020predicting} & $82.74\textsubscript{ ±0.9}$ & $75.85\textsubscript{ ±0.9}$ & $75.86\textsubscript{ ±3.3}$ & $66.01\textsubscript{ ±2.1}$ & $79.59\textsubscript{ ±1.5}$ & $69.64\textsubscript{ ±1.8}$ & $86.71\textsubscript{ ±2.3}$ & $81.79\textsubscript{ ±1.4}$ \\
SparseConvMIL \cite{lerousseau2021sparseconvmil} & $83.78\textsubscript{ ±1.4}$ & $76.54\textsubscript{ ±1.4}$ & $73.40\textsubscript{ ±3.0}$ & $68.23\textsubscript{ ±2.1}$ & $73.13\textsubscript{ ±2.4}$ & $64.77\textsubscript{ ±2.7}$ & $91.02\textsubscript{ ±0.9}$ & $82.68\textsubscript{ ±0.9}$ \\
TransMIL \cite{shao2021transmil} & $84.71\textsubscript{ ±1.8}$ & $73.88\textsubscript{ ±1.4}$ & $80.79\textsubscript{ ±1.8}$ & $69.12\textsubscript{ ±1.8}$ & $77.60\textsubscript{ ±1.8}$ & $67.70\textsubscript{ ±2.4}$ & $90.65\textsubscript{ ±1.8}$ & $84.74\textsubscript{ ±1.4}$ \\
\midrule
DTFD-MIL \cite{zhang2022dtfd} & $84.76\textsubscript{ ±2.3}$ & $79.03\textsubscript{ ±1.7}$ & $81.97\textsubscript{ ±2.3}$ & $74.79\textsubscript{ ±3.1}$ & $79.92\textsubscript{ ±3.4}$ & $71.06\textsubscript{ ±3.3}$ & $89.47\textsubscript{ ±2.3}$ & $83.69\textsubscript{ ±0.7}$ \\
\rowcolor{lightred}
Ours (DTFD-MIL) & $85.50\textsubscript{ ±1.7}$ & $80.01\textsubscript{ ±1.2}$ & $\textbf{82.13\textsubscript{ ±2.1}}$ & $\textbf{77.44\textsubscript{ ±1.9}}$ & $\textbf{81.09\textsubscript{ ±2.0}}$ & $72.22\textsubscript{ ±2.6}$ & $\textbf{92.28\textsubscript{ ±1.4}}$ & $\textbf{86.60\textsubscript{ ±0.9}}$ \\
\midrule
MHA-MIL \cite{vaswani2017attention} & $83.48\textsubscript{ ±2.0}$ & $78.87\textsubscript{ ±1.5}$ & $79.76\textsubscript{ ±3.5}$ & $75.83\textsubscript{ ±3.6}$ & $78.32\textsubscript{ ±4.4}$ & $68.80\textsubscript{ ±2.0}$ & $88.10\textsubscript{ ±0.8}$ & $84.22\textsubscript{ ±1.5}$ \\
\rowcolor{lightred}
Ours (MHA-MIL) & $\textbf{85.95\textsubscript{ ±1.4}}$ & $\textbf{80.67\textsubscript{ ±0.8}}$ & $81.34\textsubscript{ ±2.4}$ & $75.21\textsubscript{ ±1.8}$ & $80.28\textsubscript{ ±2.4}$ & $\textbf{75.01\textsubscript{ ±1.7}}$ & $92.08\textsubscript{ ±1.6}$ & $85.35\textsubscript{ ±0.9}$ \\

\bottomrule
\end{tabular}
\caption{Comparative analysis of Area Under the Curve (AUC) and Balanced Accuracy (B-Acc) scores for the GS/CIN task across CRC, STAD, and UCEC cohorts, along with the overall performance (All). The highest performance scores are highlighted in bold.}
\label{cin_compare}
\end{table*}

\section{Experiments}\label{experiments}
In our framework's experimental validation, we concentrated on two crucial cancer subtyping tasks: Microsatellite Stable (MSS) versus Microsatellite Instability (MSI), and Genomic Stability (GS) versus Chromosomal Instability (CIN). We selected these tasks due to their significant importance across various forms of cancer \cite{de2010clinical,amato2022microsatellite,vishwakarma2020chromosome,zhao2019molecular}.

\textbf{Dataset. } We utilized three cohorts from the Cancer Genome Atlas\footnote{\href{https://portal.gdc.cancer.gov}{https://portal.gdc.cancer.gov}} (TCGA) project \cite{weinstein2013cancer}: CRC (colorectal adenocarcinoma), STAD (stomach adenocarcinoma), and UCEC (endometrial carcinoma).
All slides are H\&E stained diagnostic slides, which are the reference standard for diagnostic medicine \cite{evans2018us}. Slides counts are detailed in Table \ref{slide_counts}. Labels for GS/CIN were sourced from cBioPortal\footnote{ \href{https://www.cbioportal.org}{https://www.cbioportal.org}}; labels for MSS/MSI were derived from the auxiliary materials provided by TCGA.

\textbf{Preprocessing. } Tissue regions were segmented using the Otsu method \cite{otsu1975threshold} and tessellate into 512x512 pixel patches (tiles) at a field of view (FoV) of 20x (0.5$\mu$m/px). Slides with less than 100 tissue tiles were considered curropt and were excluded from the analysis. 
Unlike previous subtyping studies \cite{bilal2021development, kather2019deep, hezi2024cimil, echle2021deep, schirris2022deepsmile}, which manually filter for tumor regions, our approach utilizes the entire slide for prediction. Although deep learning methods have achieved high accuracy in detecting tumor regions for the CRC cohort \cite{bilal2021development, chang2023predicting, jiang2022clinical}, the availability of labeled datasets for other cancer types remains limited, making it impractical for multi-cohort studies.

\textbf{Training Details. } All results were derived using 5-fold cross-validation, with folds split randomly under a fixed seed. Importantly, splits were conducted on a patient basis to prevent any leakage, and were stratified by class label and cohort association to ensure representativeness. 
Due to dataset differences, the results for all other methods were carefully reproduced using the official code under the same settings. 
We portion 10\% of the training data for validation. During training, models were validate after each epoch, and the three best models (selected based on overall patient AUC) were bagged. To further ensure robustness of the results, a few tuning steps were performed for each model, with the best result reported. In total, 16 Nvidia A100 GPUs were used.

Unless specified otherwise, a ViT-small backbone \cite{dosovitskiy2020image} was used as tile encoder. For our approach, a Cohort-Aware ViT-small backbone (CAViT) was employed, as describe in \cref{cavit}. Both encoders were pretrained under similar settings.
For additional implementation details and stability analysis, please refer to the supplementary materials.



\subsection{Benchmarking with Existing Works}
We compare our approach with several popular existing MIL models: conventional instance-level MIL methods, including Mean-Pooling and Max-Pooling; GNN-based MIL (GCN-MIL) \cite{zhao2020predicting}; convolution-based MIL (SparseConvMIL) \cite{lerousseau2021sparseconvmil}; and attention-based MIL models, such as the classic AB-MIL \cite{ilse2018attention}, TransMIL \cite{shao2021transmil}, DTFD-MIL \cite{zhang2022dtfd}, and MHA-MIL \cite{vaswani2017attention}.

For our multi-cohort approach, we selected DTFD-MIL \cite{zhang2022dtfd} and MHA-MIL \cite{vaswani2017attention} as representative models due to their consistent strong performance in our experiments. The results for MSS/MSI and GS/CIN are provided in Tables \ref{msi_compare} and \ref{cin_compare}, respectively.

For the MSS/MSI task, our approach shows significant improvements, especially in the CRC cohort, where our DTFD-MIL variant achieved the highest AUC of 87.94\%, a notable increase over baselines. In the UCEC cohort, although our approach led in both AUC and B-Acc, the overall performance was degraded. We attribute this to the frequent overlap with the POLE subtype in UCEC cases \cite{yao2024clinical}, which complicates the learning process due to its distinct characteristics. 

For the GS/CIN task, our method also delivers superior performance. Both the DTFD-MIL and MHA-MIL variants of our model lead in overall AUC and B-Acc, with the DTFD-MIL variant achieving the best results on the CRC cohort (AUC 82.13\%) and UCEC cohort (AUC 92.28\%).

In both tasks, our approach consistently performs well across all cohorts, while baseline models tend to excel in one cohort but degrade in others. This stability across cohorts highlights the effectiveness of our multi-cohort strategy in balancing cohort-specific nuances, ultimately leading to enhanced generalization and overall model performance. For more extensive stability analysis, please refer to the supplementary materials.


\subsection{Motivation for Multi-Cohort Analysis}
To further demonstrate the benefits of a multi-cohort WSI analysis, we compare our multi-cohort framework against two other training approaches: Independent Training, where each cohort is analyzed separately, and Joint Training, where all cohorts are naively combined during training. We utilize the DTFD-MIL model for this comparison, with additional analysis using the MHA-MIL model provided in the supplementary material. The results for MSS/MSI and GS/CIN tasks are shown in Tables \ref{msi_training} and \ref{cin_training}, respectively.

For the MSS/MSI task, joint training resulted in a performance drop across all cohorts, suggesting that naive joint training can be counterproductive. However, our multi-cohort approach significantly enhanced performance, exceeding both independent and joint training methods.

In the GS/CIN task, joint training improved CRC performance but had limited or negative effects on STAD and UCEC. Despite this, our approach consistently delivered superior results across all cohorts.

\begin{table}[ht]
\centering
\begin{tabular}{lccc}
\toprule
Training & CRC & STAD & UCEC \\
\midrule
Independent & 85.41 & 78.65 & 66.62 \\
Joint & 82.09 & 74.83 & 63.53 \\
\rowcolor{lightred}
Ours & \textbf{87.94} & \textbf{82.05} & \textbf{67.04} \\
\bottomrule
\end{tabular}
\caption{Performance comparison for MSS/MSI task across different training approaches.}
\label{msi_training}
\end{table}

\begin{table}[ht]
\centering
\begin{tabular}{lccc}
\toprule
Training & CRC & STAD & UCEC \\
\midrule
Independent & 73.86 & 79.67 & 91.80 \\
Joint & 81.97 & 79.92 & 89.47 \\
\rowcolor{lightred}
Ours & \textbf{82.13} & \textbf{81.09} & \textbf{92.28} \\
\bottomrule
\end{tabular}
\caption{Performance comparison for GS/CIN task across different training approaches.}
\label{cin_training}
\end{table}

\subsection{Ablation Study}
We systematically evaluate each component of our framework, highlighting its incremental contribution to overall performance. 
In this section, we focus on evaluating the overall fit of the model. Therefore, we analyze the aggregated patient AUC, rather than the individual cohort AUC as previously discussed.
As before, we selected DTFD-MIL \cite{zhang2022dtfd} and MHA-MIL \cite{vaswani2017attention} as representative models.

\subsubsection{Impact of Different Encoders} We compare our Cohort-Aware Vision Transformer (CAViT-small) with two established tile encoders: ViT-small \cite{dosovitskiy2020image} and ResNet50 \cite{he2016deep}. All encoders were pretrained under similar settings.
As presented in Table \ref{encoder_comparison}, CAViT-small consistently outperformed both ViT-small and ResNet50 across all tasks, achieving the highest AUC scores for MSS/MSI and GS/CIN with both DTFD-MIL and MHA-MIL models. Additionally, ViT-small demonstrated superior performance compared to ResNet50 in most scenarios.
    
\begin{table}[ht]
\centering
\begin{tabular}{lcccc}
\toprule
\multirow{2}{*}{Encoder} & \multicolumn{2}{c}{MSS/MSI} & \multicolumn{2}{c}{GS/CIN} \\
\cmidrule(lr){2-3} \cmidrule(lr){4-5}
 & DTFD. & MHA. & DTFD. & MHA. \\
\midrule
ResNet50 & 74.81 & 77.23 & 81.76 & 83.93 \\
ViT-small & 77.44 & 76.19 & 84.26 & 84.44 \\
\rowcolor{lightred}
CAViT-small & \textbf{79.50} & \textbf{80.60} & \textbf{85.50} & \textbf{85.95} \\
\bottomrule
\end{tabular}
\caption{Comparison of Encoder Performance (AUC) Across GS/CIN and MSS/MSI Tasks for DTFD-MIL and MHA-MIL Models}
\label{encoder_comparison}
\end{table}

\begin{figure}[ht]
\centering
\includegraphics[width=1\columnwidth]{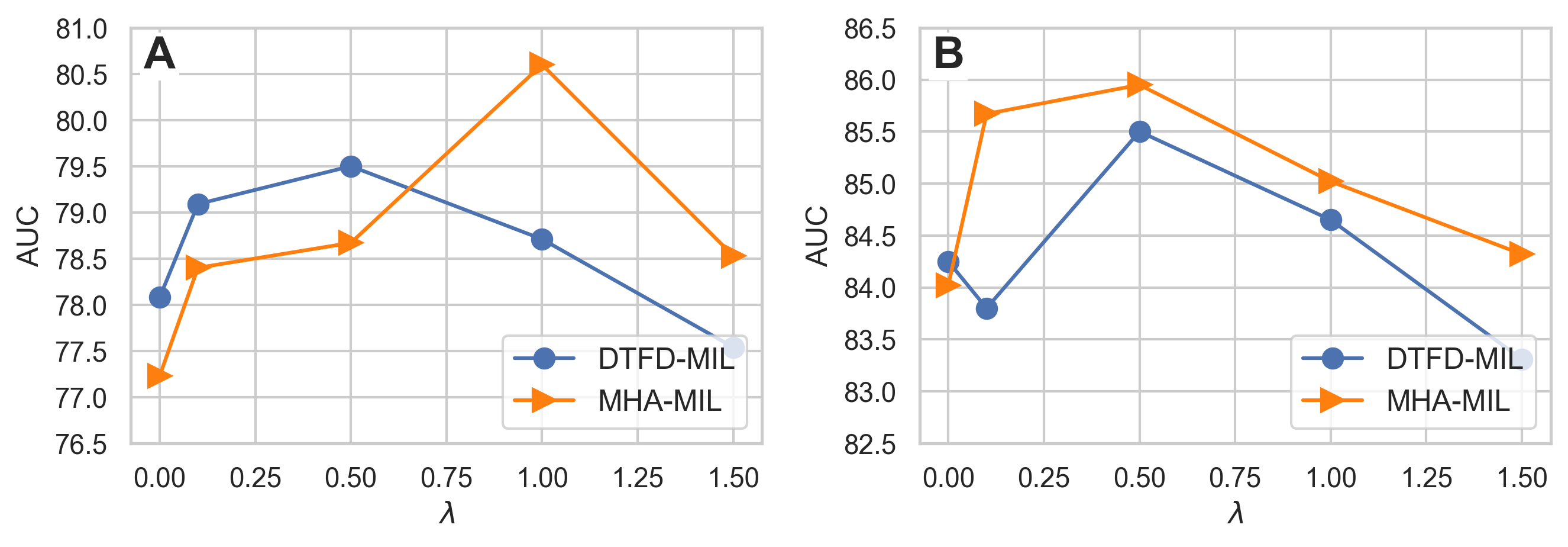}
\caption{AUC Performance with Varying \(\lambda\) Values. A: MSS/MSI Task. B: GS/CIN Task.}
\label{lambda_fig}
\end{figure}

\subsubsection{Impact of Lambda Value} 
Our approach utilizes the SMILE algorithm \cite{song2019understanding} as an adversarial estimator of mutual information (MI) between cohort associations and MIL representations. This MI estimation is incorporated as a regularization term for the main network, controlled by \(\lambda\) hyperparameter. 
As shown in \cref{lambda_fig}, higher \(\lambda\) values generally led to improved performance in both tasks, where the optimal \(\lambda\) was 1 for the MSS/MSI task and 0.5 for GS/CIN.

\subsubsection{Impact of Different Balancing Techniques} 
\begin{figure}[ht]
\centering
\includegraphics[width=0.75\columnwidth]{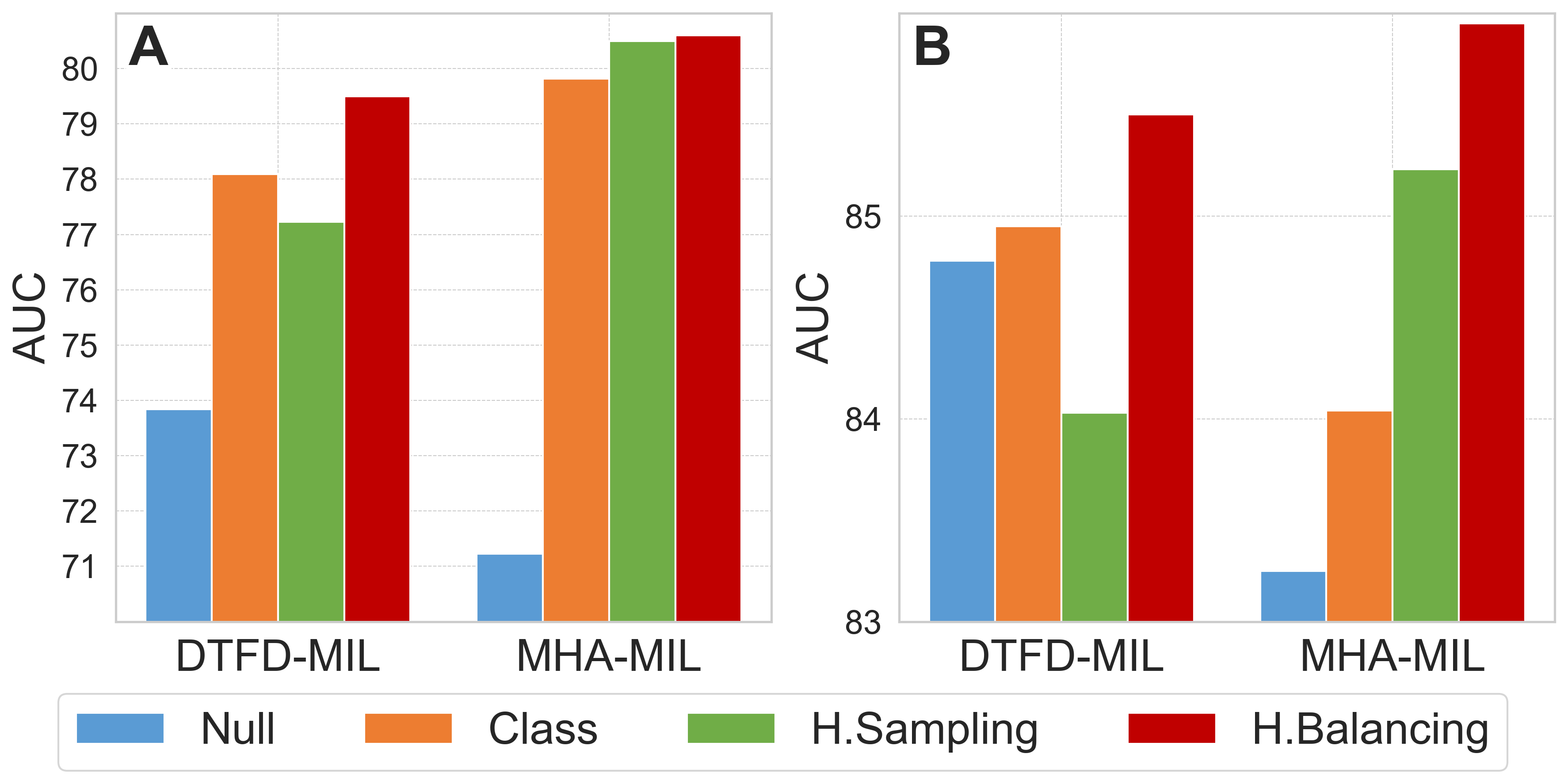}
\caption{AUC Performance of different sample weighting techniques. A: MSS/MSI Task. B: GS/CIN Task.}
\label{balancing_techniques}
\end{figure}

We evaluate our hierarchical balancing technique (H.Balancing) against other established methods: standard class balancing (Class), hierarchical weighted sampling (H.Sampling), which samples according to the hierarchical weights of samples, and when no balancing is applied (Null).
As shown in \cref{balancing_techniques}, our H.Balancing technique consistently achieved best performance across all models and tasks. For the MSS/MSI task, our approach demonstrated a notable improvement, whereas H.Sampling underperformed when used with the DTFD-MIL model.

\section{Conclusions}

Despite the availability of larger WSI datasets, combining cohorts remains essential due to the diversity and rarity of many tumor types, necessitating a multi-cohort approach. In this paper, we introduced a cohort-aware attention module integrated into a VisionTransformer for tile encoding, along with adversarial training to ensure unbiased slide representations. To address cohort imbalances, we implemented a hierarchical sample balancing strategy, enabling more balanced learning. Our results demonstrate the superiority of our approach, reinforcing the benefits of multi-cohort WSI analysis and encouraging further research into the integration of distinct cancer types, a critical step for advancing digital pathology.

{\small
\bibliographystyle{ieee_fullname}

}

\end{document}